\documentclass[12pt]{article}
\usepackage{graphics}
\def \lsim            
{\raisebox{-3pt}{$\>\stackrel{<}{\scriptstyle\sim}\>$}}
\def \gsim            
{\raisebox{-3pt}{$\>\stackrel{>}{\scriptstyle\sim}\>$}}

\begin{document}
\begin{titlepage}
\begin{flushright}
\end{flushright}
\par \vspace{10mm}
\begin{center}
{\Large \bf 
Bottom Quark Fragmentation\\[1ex]
in Top Quark Decay}
\end{center}
\par \vspace{2mm}
\begin{center}
{\bf G. Corcella and A.D. Mitov}\\
\vspace{2mm}
{Department of Physics and Astronomy, University of Rochester,\\
Rochester, NY 14627, U.S.A.}
\end{center}

\par \vspace{2mm}
\begin{center} 
{\large \bf Abstract} 
\end{center}
\begin{quote}
  \pretolerance 10000 
We study the fragmentation of the $b$ quark in top decay in NLO
QCD, within the framework of perturbative fragmentation, which
allows one to resum large logarithms $\sim\log (m_t^2/m_b^2)$.
We show the $b$-energy distribution, which we compare with
the exact ${\cal O}(\alpha_S)$ result for a massive $b$ quark.
We use data from $e^+e^-$ machines in order to describe the $b$-quark
hadronization and make predictions for the energy spectrum of $b$-flavoured
hadrons in top decay.
We also investigate the effect of NLL soft-gluon resummation
in the initial condition of the perturbative fragmentation function 
on parton- and hadron-level energy distributions.
\end{quote}

\vspace*{\fill}
\begin{flushleft}
  October 2001
\end{flushleft}
\end{titlepage}

\section{Introduction}
For the sake of performing accurate studies of the top-quark properties and a
precise measurement of its mass at the present Run II of the Tevatron 
accelerator and, in future, at the LHC [\ref{lhc}]
and at the Linear Collider [\ref{lc}],
a reliable description of the bottom-quark fragmentation in 
top decay $t\to bW$ will be essential.

As shown in [\ref{tev}], the $b$-fragmentation is indeed one of the 
sources of uncertainty in the measurement of the top mass at the 
Tevatron, as it contributes to the so-called Monte Carlo systematics.
At the LHC, recent studies [\ref{avto}] have suggested that final states 
with leptons and $J/\psi$, with the $J/\psi$ coming from the decay
of a $b$-flavoured hadron and the isolated lepton from the $W$ decay,
will be a promising channel to reconstruct the top mass.   
In [\ref{avto}], the expected experimental error,
a result of statistics and systematics, has been estimated to be 
$\Delta m_t\simeq 1$~GeV and the $b$ fragmentation is the largest source 
of uncertainty, accounting for about 0.6~GeV.

Available tools to describe the $b$-quark hadronization are
Monte Carlo event generators, such as
HERWIG [\ref{herwig}], PYTHIA [\ref{pythia}] or ISAJET [\ref{isajet}],
implementing respectively cluster [\ref{webber}], string 
[\ref{string}] and independent-fragmentation [\ref{feyn}] models.
Monte Carlo programs describe the initial- and final-state multiparton 
radiation in
hadron collisions according to 
the soft and/or collinear approximation (see, for example, 
[\ref{marweb}]). HERWIG and PYTHIA parton showers have been provided with
matrix-element corrections 
for a few processes, such as top decay [\ref{mecorr}], in order
to allow hard and large-angle parton radiation.
The analysis of [\ref{avto}] was in fact performed
using the PYTHIA event generator. In [\ref{cmlms}], the HERWIG 
event generator was used to perform studies on the top mass
reconstruction at the LHC, relying on the $b$-quark fragmentation.
The $m_{B\ell}$ invariant mass distribution, where $B$ is a 
$b$-flavoured hadron and $\ell$ a lepton from the $W$ decay,  was exploited in
order to fit the top mass. 

In this paper we analyse the 
$b$-quark fragmentation in top decay in the framework
of perturbative fragmentation at next-to-leading order (NLO) in QCD. 
The factorization theorem [\ref{collins}] 
dictates that, up to power corrections $\sim {\cal O}((1/Q)^p)$,
with $p\geq 1$ and $Q$ being a characteristic energy scale
of the process, a hadron-level cross section 
can be written as the convolution of a short-distance, perturbative 
cross section and long-distance, non-perturbative terms, 
corresponding to initial-state parton distribution functions and/or
final-state fragmentation functions.
For heavy-quark production, the quark mass $m$ acts as a regulator for the 
collinear singularity and allows one to perform perturbative calculations. 
However, fixed-order event shapes or differential distributions 
typically contain 
terms like $\alpha_S \log (Q^2/m^2)$, where $Q$ is, for example,   
the centre-of-mass energy or the 
heavy-quark transverse momentum. Such terms spoil
the convergence of the perturbative expansion and make fixed-order 
calculations unreliable once $Q$ is much larger than $m$. 
The method of perturbative fragmentation functions, originally proposed in 
[\ref{mele}], allows one to resum these large logarithms.

According to the method in [\ref{mele}], 
heavy quarks are first produced at large transverse momentum 
$m\ll p_T$, as if they were massless, and afterwards
they slow down and fragment into a massive object. 
The perturbative fragmentation function $D(\mu_F,m)$
expresses the transition of a massless parton into a massive
quark at the factorization scale $\mu_F$.

The value of $D(\mu_F,m)$ at any scale $\mu_F$ 
can be obtained by solving the 
Dokshitzer--Gribov--Lipatov--Altarelli--Parisi (DGLAP) equations 
[\ref{ap},\ref{dgl}], once its initial value at a scale $\mu_{0F}$ 
is assigned. 
The universality of the initial condition and, in general, of the perturbative 
fragmentation function, 
already suggested in [\ref{mele}] in the framework
of $e^+e^-$ annihilation, has been recently proved in a completely
process-independent way [\ref{cc}]. 
As discussed e.g.~in [\ref{cagre1}] for heavy-quark production at hadron 
colliders, the perturbative fragmentation formalism
yields a weaker dependence of phenomenological observables 
on the renormalization/factorization scales and
on the chosen set of parton distribution functions.
Furthermore, the analysis of [\ref{cc}] fully resums the leading
(LL) and next-to-leading logarithms (NLL) which are 
associated with the emission of soft gluons and appear in the 
initial condition of the perturbative fragmentation function (process
independent) and in the parton-level differential massless cross section
(process dependent) of $e^+e^-$ annihilation.

Finally, in order to describe the non-perturbative fragmentation of
a parton into a hadron, several phenomenological 
models have been proposed [\ref{kart},\ref{peterson}], besides the ones 
which are implemented in Monte Carlo event generators.
Non-perturbative fragmentation functions contain
parameters which need to be fitted to the experimental data. 
Since the hadronization mechanism is universal and independent of the 
perturbative process which produces the heavy quark, 
one can exploit
the existing data on $e^+e^-\to b\bar b$ events to fit such models 
and describe the $b$-quark
non-perturbative fragmentation in other processes, such as top decay.

Perturbative fragmentation functions and non-perturbative hadronization
models have been 
extensively applied to study the physics of $c$- and/or $b$-flavoured 
hadrons produced in
$e^+e^-$ annihilation [\ref{colnas}-\ref{nasole}], hadron collisions
[\ref{cagre1},\ref{cagre3}], Deep Inelastic Scattering [\ref{cagre2}]
and $\gamma p$ collisions [\ref{cagre4}].

In this work, we apply this method to the $b$-fragmentation in 
top decay.
In Section 2 we review the method of perturbative fragmentation
functions and apply it to predict the $b$-quark 
energy spectrum in top decay.  
In Section 3 we analyse the 
non-perturbative fragmentation of the $b$ quark and show 
energy distributions of $b$-hadrons in top decay, making use of fits
to LEP and SLD data to parametrize the hadronization models.
In Section 4 we summarize the main results of our analysis and make comments
on possible developments of our study.
In Appendices A and B we show details of our calculation and, 
comparing results for  massless and massive $b$ quarks, we 
check that our computation is consistent with the initial condition
of heavy-quark perturbative fragmentation functions.
\section{Perturbative fragmentation and parton-level results}
We wish to study $b$-quark production in
top decay within the framework of perturbative 
fragmentation functions.
We consider the decay of an on-shell top quark 
at next-to-leading order in $\alpha_S$
\footnote{Our assumption $B(t\to bW)=1$ is consistent 
with recent measurements of the CDF Collaboration of the ratio
$R=B(t\to Wb)/B(t\to Wq)$, where $q$ is a $d$, $s$ or $b$ quark, and
the subsequent 
extraction of the Cabibbo--Kobayashi--Maskawa matrix element $V_{tb}$
[\ref{vtb}].}
\begin{equation}
t(q)\to b(p_b)W(p_W)\left( g(p_g) \right)
\label{dec}
\end{equation}
\noindent 
and define the $b$-quark scaled energy fraction $x_E$ as:
\begin{equation}
x_E={{2 p_b\cdot q}\over {m_t^2}}.
\label{xbpart}
\end{equation}
Neglecting the $b$ mass, we have $0\leq x_E \leq1-w$, $w$ being 
$w=m_W^2/m_t^2$.
Throughout this paper, we shall make use of the normalized $b$ energy fraction:
\begin{equation}
x_b={{x_E}\over {1-w}}\ \ ,\ \ 0\leq x_b\leq 1.
\label{xbnorm}
\end{equation}
Following [\ref{mele}], the differential width for the production of a
massive $b$ quark in top decay can be expressed via 
the following convolution: 
\begin{equation}
{1\over {\Gamma_0}} {{d\Gamma}\over{dx_b}} (x_b,m_t,m_W,m_b)=
{1\over{\Gamma_0}}
\sum_i\int_{x_b}^1 
{{{dz}\over z}{{d\hat\Gamma_i}\over {dz}}(z,m_t,m_W,\mu,\mu_F)
D_i\left({x_b\over z},\mu_F,m_b \right)},
\label{pff}
\end{equation}
\noindent
where $\Gamma_0$ is the width of the Born process
$t\to bW$, $d\hat\Gamma_i /dz$ is the differential width for the production of
a massless parton $i$ in top decay with energy fraction $z$, 
$D_i(x_b/z,\mu_F,m_b)$ is the perturbative 
fragmentation function for a parton $i$ to fragment 
into a massive $b$ quark. In Eq.~(\ref{pff}) $\mu$ and
$\mu_F$ are the renormalization and factorization scales respectively, 
the former associated with the renormalization of the strong coupling
constant.
In principle, one
can use two different values for the factorization and renormalization
scales; however, a choice often made consists of setting
$\mu=\mu_F$ and we shall adopt this convention for most of the results
which we shall show.

The approach of [\ref{mele}] and the factorization
on the right-hand side of Eq.~(\ref{pff}) are rigorously valid if one can 
neglect
terms behaving like $(m/Q)^p$, where $p\geq 1$, 
$m$ is the mass of the fragmenting 
heavy quark and $Q$ is the hard scale of the process. This is indeed 
our case since the scale of top decay is set by its mass and
$m_b/m_t\simeq {\cal O}(10^{-2})$. 

The definitions of $d\hat\Gamma_i /dz$ and $D_i(x_b/z,\mu_F,m_b)$ 
are not unique, but they depend on the scheme which is used to subtract the
collinear singularities which appear in the 
massless differential width $d\hat\Gamma_i /dz$. 
In [\ref{mele}], the $\overline{\mathrm{MS}}$ 
factorization scheme is chosen and we shall stick to it hereinafter.

Since we have been assuming $B(t\to bW)=1$ and the probability to produce
a $b$ quark via the splitting of a secondary gluon is negligible,
we shall safely limit ourselves to considering the perturbative fragmentation
of a massless $b$ into a massive $b$ and,
on the right-hand side of Eq.~(\ref{pff}), we shall have only the
$i=b$ contribution.
We shall then need to evaluate the differential width $d\hat\Gamma_b/dz$ 
for the production of a massless $b$ quark in top decay,
in dimensional regularization, and subtract the collinear singularity 
according to the $\overline{\mathrm{MS}}$ prescription.
We obtain the following $\overline{\mathrm{MS}}$ coefficient
function:
\begin{equation}
{1\over{\Gamma_0}}{{d\hat\Gamma_b}\over {dz}}^{\overline{\mathrm{MS}}}=
\delta(1-z)+{{\alpha_S(\mu)}\over{2\pi}}\hat A_1 (z),
\label{diff}
\end{equation}
\noindent 
with
\begin{eqnarray}
\hat A_1 (z) &=& C_F\left\{
\left[ {{1+z^2}\over{(1-z)_+}}+{3\over 2}\delta (1-z)\right]
\left[\log{{m_t^2}\over {\mu_F^2}}+2 {{1+w}\over {1+2w}}-2\log (1-w)\right]
\right.
\nonumber\\
&+&{{1+z^2}\over {(1-z)_+}} \left[ 4\log \left[(1-w)z\right]-
{1\over{1+2w}}\right]
-{{4z}\over{(1-z)_+}}\left[1-{{w(1-w)(1-z)^2}\over
{(1+2w)(1-(1-w)z)}}\right]\nonumber\\
&+&2(1+z^2)\left[\left({1\over {1-z}}\log(1-z) \right)_+-{1\over {1-z}}
\log z\right]\nonumber\\
&+& \delta(1-z) \left[4{\mathrm {Li}}_2 (1-w)+
2\log(1-w)\log w -{{2\pi^2}\over 3}
+{{1+8w}\over {1+2w}}\log (1-w)\right.\nonumber\\
&-&{2w\over{1-w}}\log w+
{{3w}\over {1+2w}}-9\Bigg]\Bigg\},
\label{cms}
\end{eqnarray}
where
\begin{equation}
{\mathrm{Li}}_2 (x)=-\int_0^x {{{dt}\over t} \log (1-t)}
\end{equation}
is the Spence function.
In Appendix A we shall give more details on the derivation of
Eq.~(\ref{diff}) and present results 
for the differential width $d\Gamma/dx_b$ once the $b$-quark mass is fully 
taken into account.

In order to be consistent at NLO, 
in Eq.~(\ref{diff}) $\alpha_S(\mu)$ is to be the 
strong coupling constant at NLO
as well:
\begin{equation}
\alpha_S(\mu)={1\over {b_0\log(\mu^2/\Lambda^2)}}
\left\{ 1-{{b_1\log\left[\log (\mu^2/\Lambda^2)\right]}\over
{b_0^2\log(\mu^2/\Lambda^2)}}\right\},
\label{alpha}
\end{equation}
with $b_0$ and $b_1$ given by
\begin{equation} 
b_0={{33-2n_f}\over {12\pi}},\ \ b_1={{153-19n_f}\over{24\pi^2}},
\end{equation}
$\Lambda$ being the typical QCD scale and $n_f=5$ the number of flavours.

We note that the coefficient function (\ref{diff}) 
contains a term where the strong coupling constant
multiplies the logarithm $\log(m^2_t/\mu_F^2)$. For our 
calculation to be reliable, we shall have to require such a logarithm not
to be too large, which implies that the factorization scale $\mu_F$
will have to be chosen of the order of $m_t$.

In [\ref{mele}], considering heavy-quark production in $e^+e^-$ 
annihilation and comparing the massive and massless differential 
cross sections in the $\overline{\mathrm{MS}}$ factorization scheme, 
the authors have obtained the NLO initial conditions at a scale $\mu_{0F}$ for 
heavy-quark perturbative fragmentation functions.
For the $b\to b$ transition, it has been found
\footnote{For the fragmentation of a gluon, light quark or $\bar b$ quark into 
a $b$ quark, see [\ref{mele}].}:
\begin{equation}
D_b(x_b,\mu_{0F},m_b)=\delta(1-x_b)+{{\alpha_S(\mu_0)C_F}\over{2\pi}}
\left[{{1+x_b^2}\over{1-x_b}}\left(\log {{\mu_{0F}^2}\over{m_b^2}}-
2\log (1-x_b)-1\right)\right]_+,
\label{dbb}
\end{equation}
with $C_F=4/3$.
In order to avoid large logarithms in Eq.~(\ref{dbb}),
the scale $\mu_{0F}$ is to be taken of the order of the $b$ mass.
The universality of the initial condition (\ref{dbb})
has been lately proved in [\ref{cc}] and one can therefore
exploit it to predict the $b$ fragmentation in top decay as well.

For the sake of completeness, in Appendix A we shall show the result 
for the differential width $d\Gamma/dx_b$ once we keep the $b$ mass 
only in contributions $\sim \log(m_t^2/m_b^2)$ and neglect
terms proportional to powers of the ratio $m_b/m_t$. 
Comparing the result with the massless rate (\ref{diff}),
we shall be able to 
reproduce the initial condition of the $b$-quark fragmentation function,  
which will be a consistency check of our calculation and, at the same time, 
a confirmation of the validity of Eq.~(\ref{dbb}) in our context as well.

Assigned the initial condition (\ref{dbb}), the value of the
perturbative fragmentation function at any other scale $\mu_F$ 
can be obtained by
solving the DGLAP evolution equations
[\ref{ap},\ref{dgl}]:
\begin{equation}
{d\over{d\log\mu_F^2}}D_i(x_b,\mu_F,m_b)=
\sum_j\int_{x_b}^1{{{dz}\over z} 
P_{ij}\left({x_b\over z},\alpha_S(\mu_F)\right)
D_j(z,\mu_F,m_b)},
\label{dglap}
\end{equation}
where
\begin{equation}
P_{ij}(x_b,\alpha_S (\mu_F))={{\alpha_S(\mu_F)}\over {2\pi}}
P_{ij}^{(0)}(x_b)+\left({{\alpha_S(\mu_F)}\over {2\pi}}\right)^2
P_{ij}^{(1)}(x_b)+{\cal O} (\alpha_S^3).
\label{split}
\end{equation} 
$P_{ij}^{(0)}(x_b)$ are the Altarelli--Parisi splitting functions [\ref{ap}],
and the higher-order terms $P_{ij}^{(1)}(x_b)$ can be found in 
[\ref{pij},\ref{pij1}]. 

As shown in [\ref{mele}], solving the DGLAP equations for the evolution
from a scale $\mu_{0F}$ to $\mu_F$  
allows one to resum potentially-large logarithms 
$\sim\alpha(\mu_F)\log (\mu_F^2/\mu^2_{0F})$.
Assuming $\mu_{0F}\simeq m_b$ 
and $\mu_F\simeq m_t$, 
and considering the splitting functions (\ref{split}) at
${\cal O}(\alpha_S)$, one resums the leading logarithms
$\sim\alpha_S^n(m_t)\log^n(m_t^2/m_b^2)$. 
Accounting for ${\cal O}(\alpha_S^2)$ terms in Eq.~(\ref{split})
leads to the inclusion of next-to-leading logarithms 
$\sim\alpha_S^{n+1}(m_t)\log^n(m_t^2/m_b^2)$ as well.
In this paper, we shall always assume that the 
$b$-quark perturbative fragmentation function evolves with NLL accuracy.

The DGLAP equations get highly simplified in Mellin space;
the solution reads [\ref{mele}]:
\begin{eqnarray}
D_{i,N}(\mu_F,m_b)&=&
D_{i,N}(\mu_{0F},m_b)\exp\Bigg\{P_N^{(0)}t\nonumber\\
&+&{1\over{4\pi^2b_0}}
\left[ \alpha_S(\mu_{0F})-\alpha_S(\mu_F) \right]
\left[P_N^{(1)}-{{2\pi b_1}\over {b_0}}P_N^{(0)}\right]\Bigg\},
\label{dresum}
\end{eqnarray}
with
$D_{i,N}(\mu_F,m_b)$ being the Mellin transform of the $x$-space perturbative
fragmentation function
\begin{equation}
D_{i,N}(\mu_F,m_b)=\int_0^1{dx\  x_b^{N-1} D_i(x_b,\mu_F,m_b)}
\label{mell}
\end{equation}
and the variable $t$ defined as
\begin{equation}
t={1\over{2\pi b_0}}\log {{\alpha_S(\mu_{0F})}\over{\alpha_S(\mu_F)}}.
\end{equation}
Throughout our analysis, $i=b$ in Eqs.~(\ref{dglap}-\ref{mell})  and 
$D_{b,N}(\mu_{0F},m_b)$ is the $N$-space transform of Eq.~(\ref{dbb}).
Expressions for $D_{b,N}(\mu_{0F},m_b)$ 
and for 
the Mellin transforms of the NLO splitting functions in Eq.~(\ref{split})
$P_N^{(0)}$ and $P_N^{(1)}$ can be found in [\ref{mele}].
We shall have to compute the $N$-space 
transform of the $\overline{\mathrm{MS}}$
coefficient function (\ref{diff})
\footnote{In Appendix B we shall present results for 
the $N$-space counterpart of Eq.~(\ref{diff}).} and multiply it
by Eq.~(\ref{dresum}) in order to get 
the Mellin transform of Eq.~(\ref{pff}):
\begin{equation}
\Gamma_N(m_t,m_W,m_b)=\hat\Gamma_N(m_t,m_W,\mu,\mu_F)
D_{b,N}(\mu_F,m_b),
\label{gamman}
\end{equation}
with 
\begin{equation}
\Gamma_N (m_t,m_W,m_b)=
{1\over{\Gamma_0}}\int_0^1 {dx_b \ x_b^{N-1}
{{d\Gamma}\over{dx_b}}(x_b,m_t,m_W,m_b) }.
\end{equation}
The $b$-quark energy distribution in 
$x$-space will finally 
be obtained by inverting the $N$-space result (\ref{gamman}) numerically.

Furthermore, in Eq.~(\ref{dbb})
the coefficient multiplying the strong coupling constant contains 
terms behaving $\sim 1/(1-x_b)_+$ or $\sim [\log (1-x_b)/(1-x_b)]_+$
once $x_b\to 1$, which corresponds to behaviours 
$\sim\log N$ or $\sim\log^2 N$ in moment space, for large $N$.
The limit $x_b\to 1$ ($N\to\infty$) corresponds to soft-gluon radiation in
top decay.
Soft LL $\sim \alpha_S^n(\mu_0)\log^{n+1}N$
and NLL $\sim \alpha_S^n(\mu_0)\log^nN$
contributions in the initial condition of the
perturbative fragmentation function have been resummed in [\ref{cc}].
Due to the process independence of the heavy-quark 
perturbative fragmentation function, we can exploit the result of [\ref{cc}],
which we do not report here for the sake of brevity,
to resum LL and NLL terms in Eq.~(\ref{dresum}) 
in top decay as well.

We wish to present results for the normalized 
$b$-quark energy distribution in top decay
using the technique just described. We normalize our plots to the total 
NLO width $\Gamma$, obtained neglecting powers $\sim (m_b/m_t)^p$, 
whose expression can be found in
[\ref{wid1}]. In fact, the factorization on the right-hand side of 
Eq.~(\ref{pff}), the 
DGLAP evolution 
and the resummation of soft logarithms in the initial condition of
the perturbative fragmentation function with NLL accuracy 
do not affect the total NLO normalization\footnote{As far as soft-gluon
resummation is concerned, matching the resummed 
initial condition to the exact ${\cal O}(\alpha_S)$ result (\ref{dbb})
(see Eq.~(76) in Ref.[\ref{cc}]), guarantees that the total
NLO width is left unchanged.}. 

It can be observed that, as long as one neglects interference between top 
production and decay, one has:
\begin{equation}
{1\over{\sigma}}{{d\sigma}\over {dx_b}}={1\over {\Gamma}}
{{d\Gamma}\over{dx_b}},
\end{equation}
where $(1/\sigma) d\sigma/dx_b$ is the normalized differential cross section
for the production of a $b$ quark with energy fraction $x_b$ via top quarks,
independently of the production
mechanism. Our results will then be applicable to $p\bar p$ (Tevatron),
$pp$ (LHC) or $e^+e^-$ (Linear Collider) collisions.
For our numerical study, 
we shall assume $m_t=175$~GeV, $m_W=80$~GeV, $m_b=5$~GeV and 
$\Lambda=200$~MeV.

\begin{figure}
\centerline{\resizebox{0.65\textwidth}{!}{\includegraphics{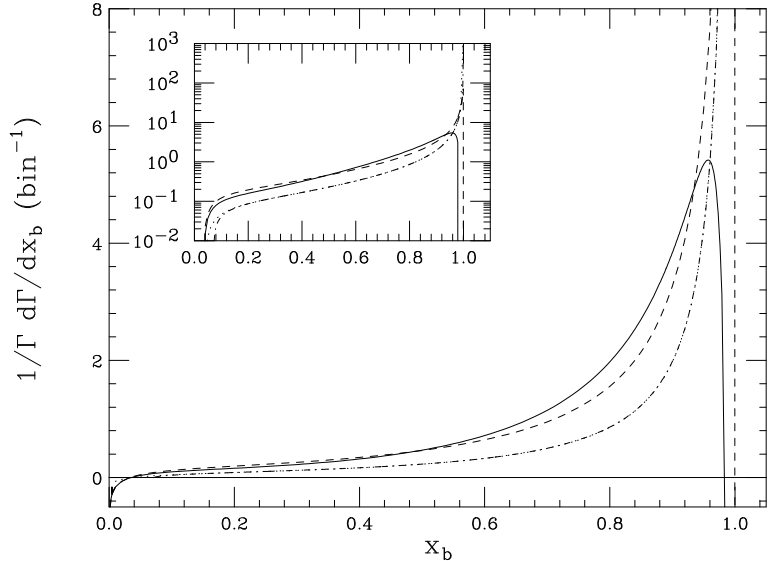}}}
\caption{$b$-quark energy distribution in top decay according to 
the perturbative fragmentation approach, with (solid line) and without 
(dashes) NLL soft-gluon resummation in the initial condition of
the perturbative fragmentation function, and according to
the exact NLO calculation, with (dot-dashes) and without (dots) 
inclusion of powers of $m_b/m_t$. In the inset figure, we show the
same curves on a logarithmic scale.}
\label{figpart}
\end{figure}

We show our results for the $x_b$ spectrum in Fig.~\ref{figpart}.
We plot the $x_b$ 
distribution according to the perturbative fragmentation approach,
with and without NLL
soft-gluon resummation in the initial condition of the perturbative
fragmentation function.
For the sake of comparison, we also show the exact 
${\cal O}(\alpha_S)$ result
for a massive $b$ quark, whose analytical expression will
be given in Appendix A. We set  $\mu=\mu_F=m_t$ and $\mu_0=\mu_{0F}=m_b$. 

We note that the use of perturbative fragmentation functions has a stronger 
impact on the $x_b$ distribution. The fixed-order result lies well below
the perturbative fragmentation 
results for about $0.1\lsim x_b\lsim 0.9$ and diverges once 
$x_b\to 1$, due to a behaviour $\sim 1/(1-x_b)_+$. Moreover,
the full inclusion of powers of $m_b/m_t$ has
a negligible effect on the $x_b$ spectrum; the dot-dashed and 
dotted lines in Fig.~\ref{figpart} are in fact almost indistinguishable.
As for the perturbative fragmentation results, 
the distribution with no soft-gluon resummation shows a very sharp 
peak, though finite, once $x_b$ approaches unity.
This behaviour is smoothed out after we resum the soft NLL logarithms appearing
in the initial condition of the perturbative fragmentation function, 
as the $b$-energy spectrum gets softer and shows the
so-called `Sudakov peak'. Both perturbative fragmentation 
distributions become negative for $x_b\to 0$ and $x_b\to 1$,
which is a known result, already found for heavy-quark
production in $e^+e^-$ annihilation [\ref{cc},\ref{nasole}].
For $x_b\to 0$, the coefficient function (\ref{diff}) contains
large logarithms $\sim\alpha_S(\mu) \log x_b$ 
which have not been resummed yet.
Likewise, in the soft limit $x_b\to 1$, Eq.~(\ref{cms}) contains contributions
$\sim\alpha_S(\mu)/(1-x_b)_+$ and 
$\sim\alpha_S(\mu)[\log(1-x_b)/(1-x_b)]_+$.
Since $\alpha_S(m_b)\simeq 2\alpha_s(m_t)$,
for $\mu= m_t$ and $\mu_0=m_b$ such terms are smaller than the similar 
ones which appear in the initial condition of the perturbative fragmentation 
function (\ref{dbb}), 
but nonetheless they would need to be resummed.
As stated in [\ref{cc}], once $x_b$ gets closer to unity, 
non-perturbative contributions also 
become important and should be taken into account. The region of 
reliability of the perturbative calculation at large $x_b$ may be related 
to the Landau pole in the expression for the strong coupling constant 
(\ref{alpha}), and estimated to be $x_b\lsim 1-\Lambda/m_b$. 

\begin{figure}
\centerline{\resizebox{0.49\textwidth}{!}{\includegraphics{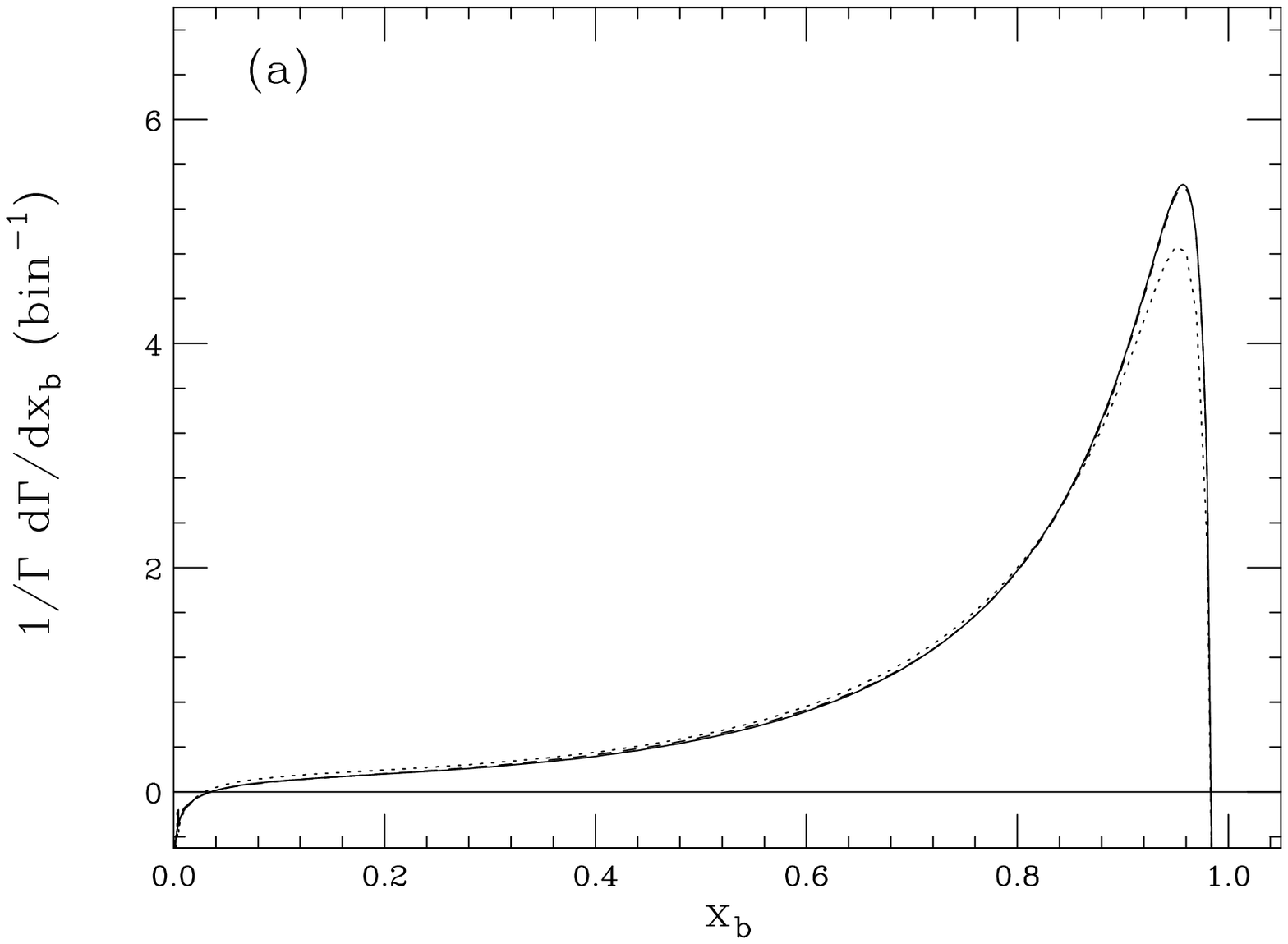}}%
\hfill%
\resizebox{0.49\textwidth}{!}{\includegraphics{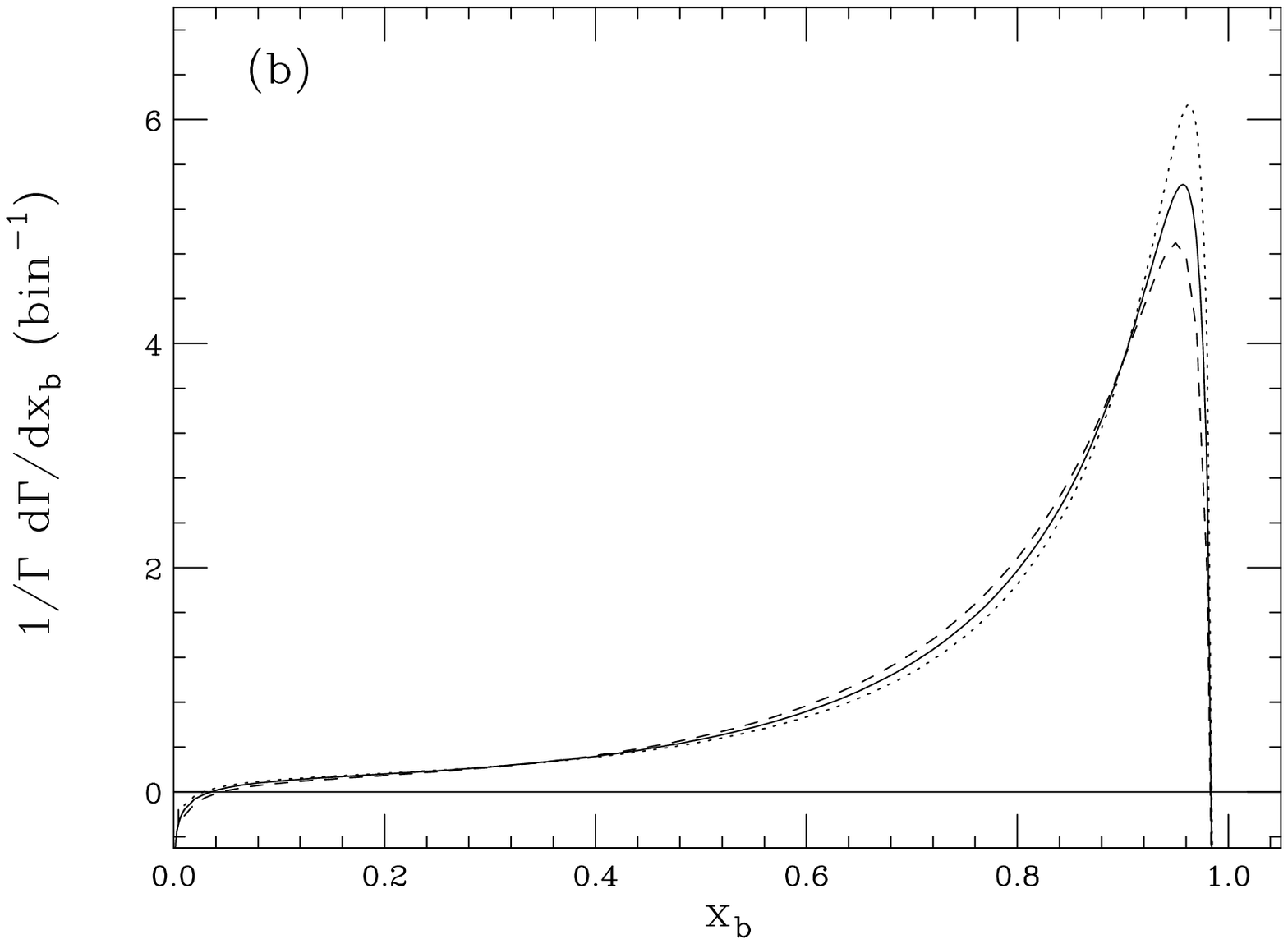}}}
\caption{(a): $x_b$ spectrum for $\mu_F = m_t$ and 
$\mu_{0F}=m_b/2$ (dots), $\mu_{0F}=m_b$ (solid) and
$\mu_{0F}=2m_b$ (dashes); (b): $\mu_{0F}=m_b$ and $\mu_F=m_t/2$ (dots), 
$\mu_F=m_t$ (solid) and 
$\mu_F=2m_t$ (dashes). 
The renormalization scales are kept at $\mu=m_t$ and $\mu_0=m_b$.
All curves include NLL soft-gluon resummation in the initial condition of
the perturbative fragmentation function.}
\label{figmu}
\end{figure}

\begin{figure}
\centerline{\resizebox{0.49\textwidth}{!}{\includegraphics{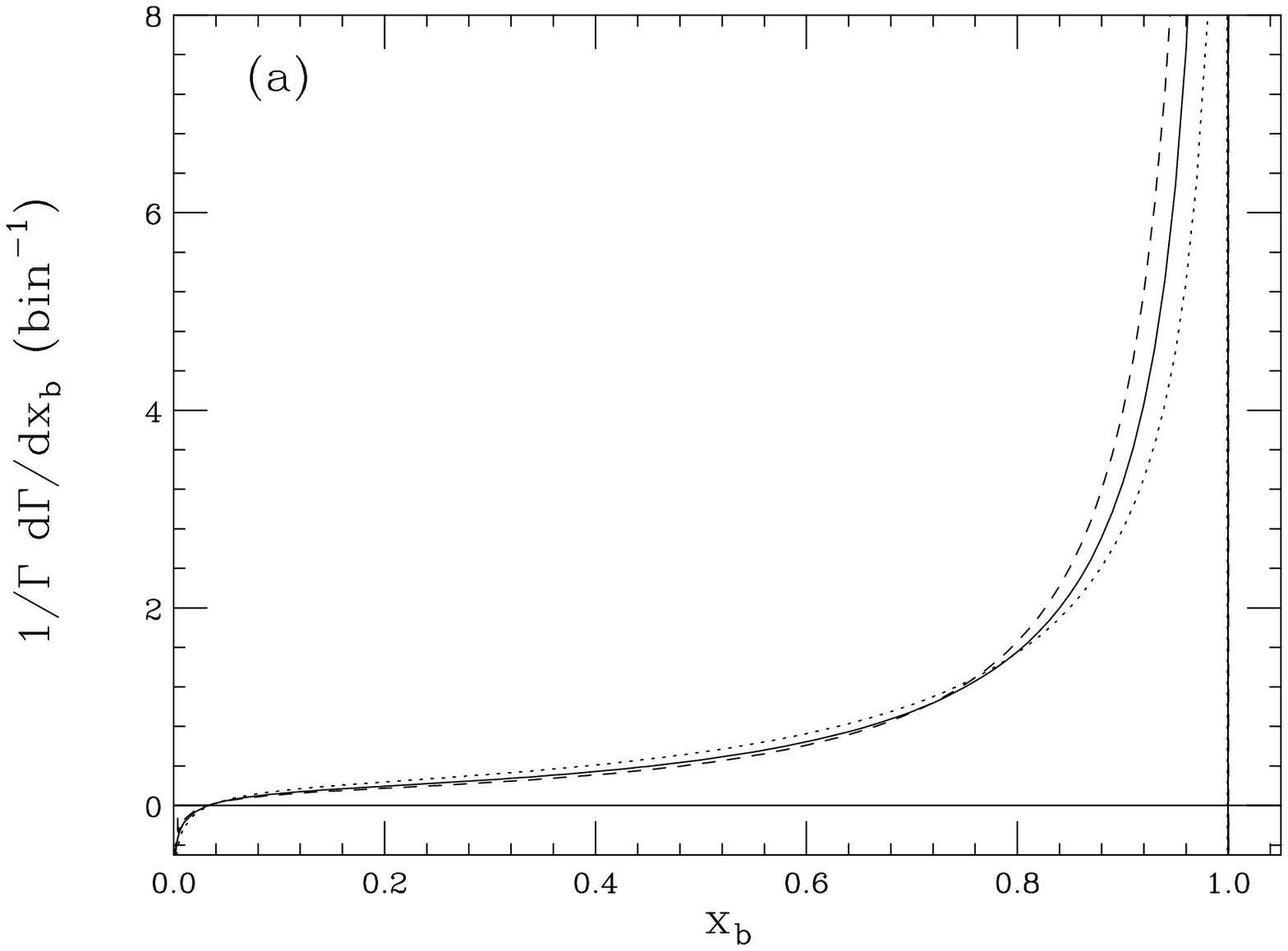}}%
\hfill%
\resizebox{0.49\textwidth}{!}{\includegraphics{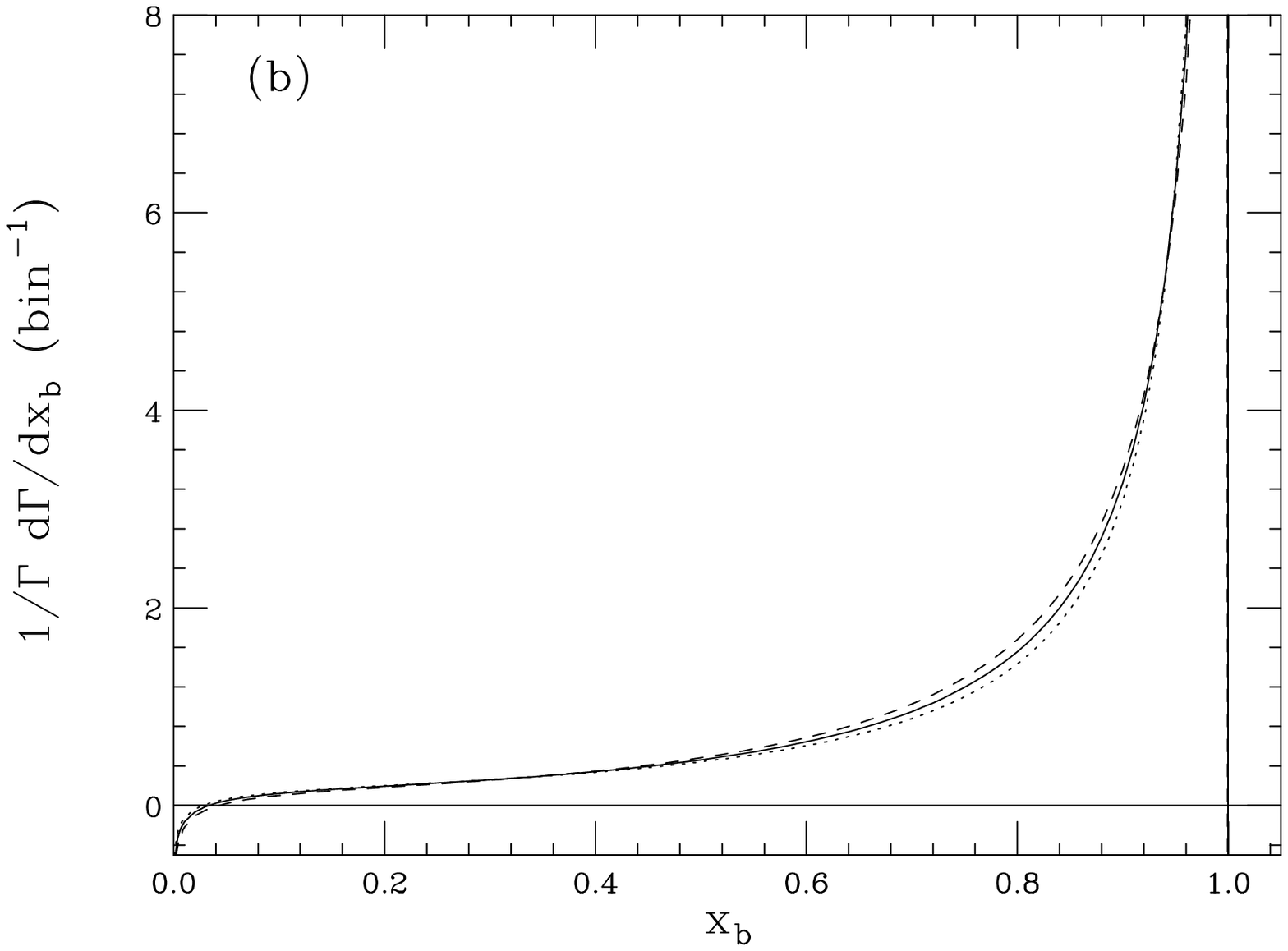}}}
\caption{As in Fig.~\ref{figmu}, but with no soft resummation.
Though not visible, all distributions show a finite, sharp peak
once $x_b$ is close to 1.}
\label{figmunos}
\end{figure}

Figs.~\ref{figmu} and \ref{figmunos} show the dependence of the 
perturbative fragmentation $x_b$ 
distributions on the factorization scales $\mu_{0F}$ and $\mu_F$, 
with (Fig.~\ref{figmu})
or without (Fig.~\ref{figmunos}) soft-gluon resummation.
We observe in Fig.~\ref{figmu}~(a) that the
dependence on the initial scale $\mu_{0F}$ is small 
when we resum soft logarithms in the initial condition of the
$b$-quark perturbative fragmentation function. Little impact is only visible
in the neighbourhood of the Sudakov peak and, in particular, 
the distributions obtained for $\mu_{0F}=m_b$ and $\mu_{0F}=2m_b$ are
indistinguishable from each other.
On the contrary, 
Fig.~\ref{figmunos}~(a) shows that the $x_b$ spectrum has a remarkable 
dependence on $\mu_{0F}$ once 
soft logarithms are not resummed. Comparing Fig.~\ref{figmu}~(b) 
and Fig.~\ref{figmunos}~(b), we note a quite similar dependence on the
choice of $\mu_F$. In fact, $\mu_{0F}$ rather than
$\mu_F$ is the scale which enters the expression of the initial 
condition $D_b(x_b,\mu_{0F},m_b)$. 
For $\mu_F$ approaching $\mu_{0F}$, the distribution with no soft 
resummation gets closer to
the fixed-order, unevolved one shown in Fig.~\ref{figpart}.
We expect that once one includes soft resummation in the 
$\overline{\mathrm{MS}}$ coefficient function (\ref{diff}), which contains
the factorization scale $\mu_F$, the dependence
of the $b$-energy spectrum on $\mu_F$
will become weaker as well, as found in [\ref{cc}] for the purpose
of $e^+e^-$ annihilation. As a whole, we can state that 
resumming soft logarithms
yields a reduction of the theoretical uncertainty, 
as the dependence on factorization scales is indeed an estimation of 
effects of higher-order contributions which we have been neglecting.
\section{Non-perturbative fragmentation and hadron-level results}
In this Section we shall present results for the energy distribution of
$b$-flavoured hadrons in top decay. We consider the transition
$b\to B$, where $B$ is either a meson or a baryon containing a $b$ quark
and define the normalized $b$-hadron energy fraction $x_B$ similarly to
the parton-level one in Eq.~(\ref{xbnorm}).

The energy distribution of a hadron $B$ can be expressed 
as the convolution of the parton-level spectrum
with the non-perturbative fragmentation function
$D_{np}^B (x_B)$:
\begin{equation}
{1\over {\Gamma_0}} {{d\Gamma}\over{dx_B}} (x_B,m_t,m_W,m_b)={1\over{\Gamma_0}}
\int_{x_B}^1 {{{dz}\over z}{{d\Gamma}\over {dz}}(z,m_t,m_W,m_b)
D_{np}^B\left({x_B\over z}\right)}.
\label{npff}
\end{equation}
In Eq.~(\ref{npff}), $(1/\Gamma_0) d\Gamma/dz$ is the parton-level differential
width (\ref{pff}) for $x_b=z$ and 
$D_{np}^B(x_B/z)$ is the non-perturbative fragmentation function describing
the hadronization $b\to B$, which is process independent.
We can therefore exploit data from 
$e^+e^-\to b\bar b$ processes to predict the
$b$-quark hadronization in top decay.

Several models have been proposed to describe the non-perturbative
transition from a quark- to a hadron-state.
One of the most-commonly used 
[\ref{colnas},\ref{cagre2},\ref{cagre3},\ref{cagre4}]
consists of a simple power functional form:
\begin{equation}
D_{np}(x;\alpha,\beta)={1\over{B(\beta +1,\alpha +1)}}(1-x)^\alpha x^\beta,
\label{ab}
\end{equation}
$B(x,y)$ being the Euler Beta function. 

The model of Kartvelishvili et al. [\ref{kart}] is still a power law,  
but with just one free parameter $\delta$:
\begin{equation}
D_{np}(x;\delta)={1\over {(1+\delta)(2+\delta)}} (1-x) x^\delta.
\label{kk}
\end{equation}           
We expect that if we use the
non-perturbative fragmentation function in (\ref{ab}), which has two 
fittable parameters, we shall be able to get better 
agreement to the data than when using (\ref{kk}). 
However, in our analysis we shall try to tune the model (\ref{kk}) as well,
in order to investigate how good it is at reproducing
the data and how it compares to the other models considered.
    
Finally, the Peterson model [\ref{peterson}] describes the transition
of a heavy quark into a heavy hadron according to the following 
non-perturbative fragmentation function:
\begin{equation}
D_{np}(x;\epsilon)={A\over {x[1-1/x-\epsilon/(1-x)]}}.
\label{peter}
\end{equation}
For an explicit expression of the normalization factor $A$ and
of the Mellin transform of Eq.~(\ref{peter}), see [\ref{spira}].
The $N$-space transforms of Eqs.~(\ref{ab}) and (\ref{kk}) are 
straightforward. 
The parameters $\alpha$ and $\beta$ in Eq.~(\ref{ab}), $\delta$ in
Eq.~(\ref{kk}) and $\epsilon$
in Eq.~(\ref{peter}) will have to be obtained from fits to  
experimental data.

To predict the $b$-quark hadronization properties, we shall
use LEP data from the 
ALEPH Collaboration [\ref{aleph}] and 
SLD data [\ref{sld}] for $e^+e^-$ collisions at the $Z$ pole, i.e.
$\sqrt{s}=91.2$~GeV. 
Both data sets refer 
to weakly-decaying $b$-hadrons \footnote{By weakly-decaying
$b$-hadrons we mean hadrons containing a $b$ quark which decay according
to a weak transition. For example, in the decay chain 
$B^*\to B\gamma\to \left(D^{(*)}\ell\nu\right) \gamma$, the $B$ meson
rather than the $B^*$ is 
the weakly-decaying $b$-hadron whose energy fraction is 
experimentally measurable.}, however, while the ALEPH $x_B$-data just  
accounts for $B$ mesons, the SLD set also considers $b$-flavoured baryons, 
mainly the $\Lambda_b$. In [\ref{aleph}], it is stated that the mean values
$\langle x_B\rangle$ 
of ALEPH and SLD are consistent with each other, within the range of systematic
and statistical errors. However, no complete
statistical analysis aiming at checking the 
consistency of the $x_B$ distributions has been done yet; therefore
some difference is to 
be expected when comparing fits of non-perturbative models
to data actually referring to different $b$-hadron samples, as we do have.

In order for the results of our fits to be applicable to the $b$-hadronization
in top decay, we shall have to describe the perturbative 
process $e^+e^-\to b\bar b (g)$ in the same framework as we did for
$t\to bW(g)$ when we do the fits. 
In fact, the factorization on the right-hand side of Eq.~(\ref{npff})
and the splitting between perturbative and non-perturbative part is
somewhat arbitrary and the parametrization of the non-perturbative 
model indeed depends on the approach which is 
used to describe the perturbative,
parton-level process and on the values which are chosen for quantities  
like $\Lambda$, $m_b$ and for the renormalization and
factorization scales.
We shall therefore use $\overline{\mathrm{MS}}$
coefficient functions [\ref{nasweb}] 
for $e^+e^-$ annihilation into massless quarks
and convolute them with the perturbative $b$-quark fragmentation
function evolved to NLL accuracy according to the DGLAP equations,
possibly including soft-gluon NLL resummation in the initial condition of
the perturbative fragmentation function.
We shall then obtain a parton-level differential cross section 
$1/\sigma_0 (d\sigma/dx_b)$, equivalent to Eq.~(\ref{pff}), 
which we shall have to convolute with the
non-perturbative fragmentation function of the hadronization model
considered.

The experimental analyses [\ref{aleph},\ref{sld}] use 
the PYTHIA Monte Carlo event generator 
[\ref{pythia}] to simulate $e^+e^-\to b\bar b$ 
processes and the subsequent parton
showering; as a result, their 
framework is pretty different from the one which we have 
been using for top decay.
We cannot therefore simply use
the parametrizations of the hadronization models as they are reported  
in [\ref{aleph},\ref{sld}].

An extensive phenomenological study of fragmentation
in $e^+e^-$ processes, with more details on fits to LEP
and SLD data is currently under way [\ref{caccia}].
For the purposes of our paper, we just  
tune the hadronization models to the data sets
for a particular choice of the quantities
which enter the perturbative calculation. 

We point out that, although we convolute our perturbative result with a
non-perturbative, smooth function, problems are still to be expected once 
$x_B\to 0,1$. 
The region $x_B\to 0$ will not be reliably described since 
the perturbative calculation itself includes
unresummed $\sim \log x_b$ terms.
For $x_B\to 1$ one should in principle correctly 
account for all the missing, non-perturbative power corrections.
Resumming a class of perturbative soft logarithms and
using a specific functional form with few fittable parameters  
to describe the non-perturbative fragmentation 
is not sufficient to be able to include all such terms.

In order to perform trustworthy fits to the $e^+e^-$ data and acceptable
predictions for the $b$-hadron spectrum in top decay, we shall limit 
ourselves to analysing data not too close to the critical points
$x_B=0,1$. In particular, we shall consider ALEPH data within the
range $0.18\lsim x_B \lsim 0.94$ and SLD data for $0.18\lsim x_B\lsim 0.90$,
which implies that our predictions for top decay will have to
be considered in the same $x_B$ ranges. When doing the fits, 
we account for both statistical and systematic errors on the data.

In Tables~\ref{tableres} and \ref{tablenores} we report the parameters 
which correspond to 
our best fits to the data, along with the $\chi^2$ per
degree of freedom. We also investigate the impact
of NLL soft-gluon resummation in the initial condition
of the perturbative fragmentation
function. Standard deviations for best-fit parameters are included as well.
\begin{table}[t]
\begin{center}
\begin{tabular}{||l|r|r||}\hline
&ALEPH\hspace{1.05cm} &SLD\hspace{1.45cm} \\ \hline 
\hspace{1.cm}$\alpha$&$0.31\pm 0.15$\hspace{0.79cm} &$1.88\pm 0.42$
\hspace{0.78cm} \\ \hline
\hspace{1.cm}$\beta$&$13.21\pm 1.62$\hspace{0.8cm} 
&$27.04\pm 4.02$\hspace{0.92cm} \\ \hline
$\chi^2(\alpha,\beta)$/dof&2.62/14\hspace{1.2cm} 
&11.12/16\hspace{1.2cm} \\ \hline
\hspace{1.cm}$\delta$&$20.39\pm 0.77$\hspace{0.8cm} 
&$18.80\pm 0.60$\hspace{0.92cm} \\ \hline
\hspace{0.2cm}$\chi^2(\delta)$/dof&17.27/15\hspace{1.2cm} 
&17.46/17\hspace{1.1cm}  \\ \hline
\hspace{1.cm}$\epsilon$&$(1.12\pm 0.16)\times 10^{-3}$
&$(1.17\pm 0.10)\times 10^{-3}$\\ \hline
\hspace{0.2cm}$\chi^2(\epsilon)$/dof&22.96/15\hspace{1.1cm}
&130.80/17\hspace{1.1cm} \\ \hline
\end{tabular}
\end{center}
\caption{Results of fits to $e^+e^-\to b\bar b$ data, using NLO   
coefficient functions, NLL DGLAP evolution and NLL soft-gluon
resummation in the initial condition of the 
perturbative fragmentation function. 
We set $\Lambda=200$~MeV, $\mu_{0F}=\mu_0=m_b=5$~GeV and
$\mu_F=\mu=\sqrt{s}=91.2$~GeV. $\alpha$ and $\beta$ are the parameters in 
the power law (\ref{ab}), $\delta$ refers to (\ref{kk}), $\epsilon$ 
to (\ref{peter}).
\label{tableres}}
\end{table}
\begin{table}[t]
\begin{center}
\begin{tabular}{||l|r|r||}\hline
&ALEPH\hspace{1.2cm} &SLD\hspace{1.45cm} \\ \hline 
\hspace{1.cm}$\alpha$&$0.66\pm 0.13$\hspace{0.9cm} 
&$2.05\pm 0.28$\hspace{0.9cm} \\ \hline
\hspace{1.cm}$\beta$&$12.39\pm 1.04$\hspace{0.9cm}
&$22.10\pm 2.13$\hspace{0.9cm} \\ \hline
$\chi^2(\alpha,\beta)$/dof&7.12/14\hspace{1.3cm} 
&40.23/16\hspace{1.2cm} \\ \hline
\hspace{1.cm}$\delta$&$14.97\pm 0.44\hspace{0.9cm} 
$&$14.57\pm 0.37$\hspace{0.9cm} \\ \hline
\hspace{0.2cm}$\chi^2(\delta)$/dof&13.30/15 \hspace{1.1cm} 
&58.63/17\hspace{1.1cm} \\ \hline
\hspace{1.cm}$\epsilon$&$(2.87\pm 0.21)\times 10^{-3}$
&$(2.33\pm 0.19)\times 10^{-3}$\\ \hline
\hspace{0.2cm}$\chi^2(\epsilon)$/dof&52.76/15\hspace{1.2cm} 
&275.69/17\hspace{1.1cm} \\ \hline
\end{tabular}
\end{center}
\caption{As in Table~\ref{tableres}, but without resumming soft
logarithms.
\label{tablenores}}
\end{table}
\par
From Table~\ref{tableres}, we 
learn that the use of the power law (\ref{ab}) which has two 
tunable parameters leads to excellent fits
to both ALEPH and SLD data, but the values of $\alpha$ and $\beta$ 
which minimize the $\chi^2$ are affected by fairly large errors.
The model of Kartvelishvili et al. is good at fitting in with the 
ALEPH and SLD data as well. 
The Peterson model is marginally consistent with the ALEPH data and 
unable to reproduce the SLD data.
Although we are comparing data samples with different $b$-hadron contents, 
we observe that the best-fit values of $\epsilon$ and $\delta$ obtained
for ALEPH and SLD are compatible within one and two standard 
deviations respectively. A bigger difference between ALEPH and SLD 
is found once we try to fit the power law (\ref{ab}). 

Comparing Tables~\ref{tableres} and \ref{tablenores} we observe that the  
implementation of soft-gluon resummation in the 
perturbative fragmentation function results in statistically-different 
values of the best-fit parameters.
With no soft resummation, using power laws with one or two 
fittable parameters still yields very-good 
fits to the ALEPH data, while none of the models considered
is able to describe consistently the SLD data and the Peterson
non-perturbative fragmentation fails in reproducing the ALEPH data as well.

We wish now to predict the $b$-hadron spectra in top decay, using models
and parameters which give reliable fits to the $e^+e^-$ data. 
In order to account for the uncertainties on the parameters 
in the non-perturbative fragmentation functions, as  
shown in Tables~\ref{tableres} and \ref{tablenores},
for each hadronization model we shall plot
two curves which delimit a set of predictions at one-standard-deviation
confidence level
\footnote{We point out that 
the correlations between the errors on $\alpha$ and $\beta$ of 
the non-perturbative fragmentation function (\ref{ab}) are correctly 
taken into account throughout our analysis and in the plots which we
show.}. 

\begin{figure}
\centerline{\resizebox{0.60\textwidth}{!}{\includegraphics{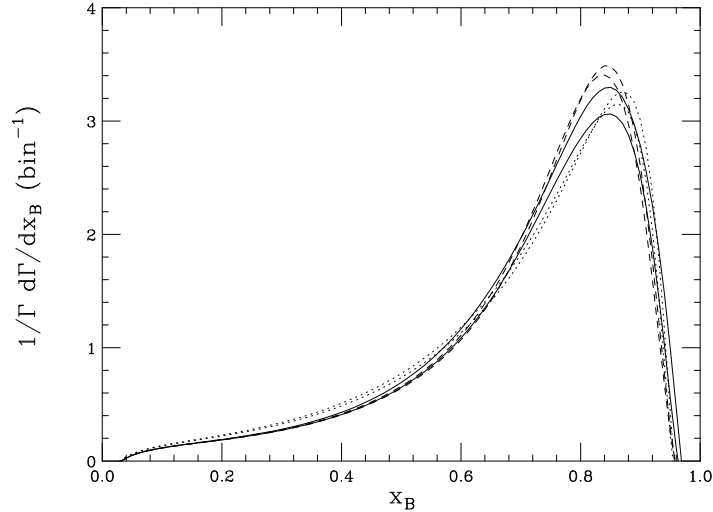}}}
\caption{$x_B$ spectrum in top decay, with the hadronization 
modelled according to a power law (solid lines), 
the Kartvelishvili et al. (dashes) and the Peterson (dots) model,
with the relevant parameters fitted to the ALEPH 
data. The plotted curves are the edges of bands at
one-standard-deviation confidence level. NLL soft-gluon resummation
in the initial condition of the 
perturbative fragmentation function is included.
We set $\mu_F=\mu=m_t$ and $\mu_{0F}=\mu_0=m_b$.}
\label{had1}
\end{figure}

\begin{figure}
\centerline{\resizebox{0.60\textwidth}{!}{\includegraphics{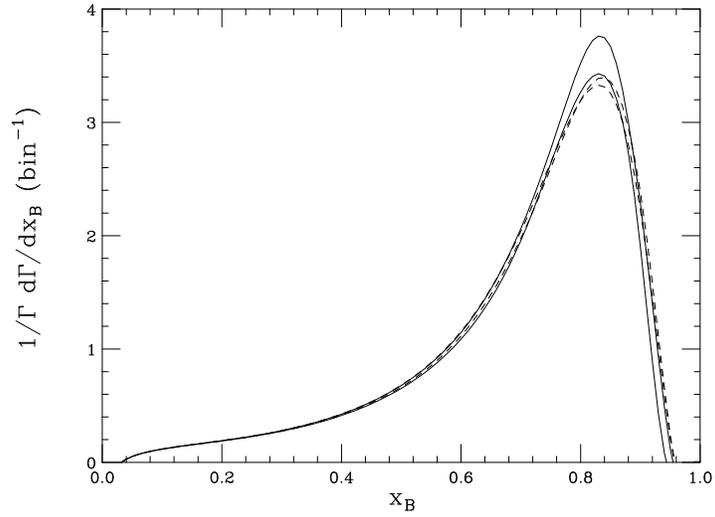}}}
\caption{As in Fig.~\ref{had1}, but fitting the hadronization-model
parameters to the SLD data.}
\label{had2}
\end{figure}

In Figs.~\ref{had1} and 
\ref{had2} we show the $x_B$ distribution using all three
considered models fitted to the ALEPH data (Fig.~\ref{had1}) 
and the power forms 
(\ref{ab}) and (\ref{kk}) fitted to SLD (Fig.~\ref{had2}), with all 
curves including NLL soft resummation in the initial condition of the
perturbative fragmentation 
function. 
We note that different hadronization models yield statistically-different 
predictions for $b$-hadron spectra in top decay, within the accuracy
of one standard deviation.
However, one can show that the $x_B$ distributions according to
the models (\ref{ab}) and (\ref{kk}) fitted to the SLD data
are consistent within two standard deviations.
This result can be related 
to the use of similar functional forms and to the large
errors on $\alpha$ and $\beta$. The prediction obtained fitting the Peterson
non-perturbative fragmentation function to the ALEPH data
looks pretty different from the others, especially   
at small and middle values of $x_B$, where the predicted 
errors are pretty small for all the considered models.
Moreover, the Peterson distributions are peaked at slightly-larger 
values of $x_B$. 

\begin{figure}
\centerline{\resizebox{0.60\textwidth}{!}{\includegraphics{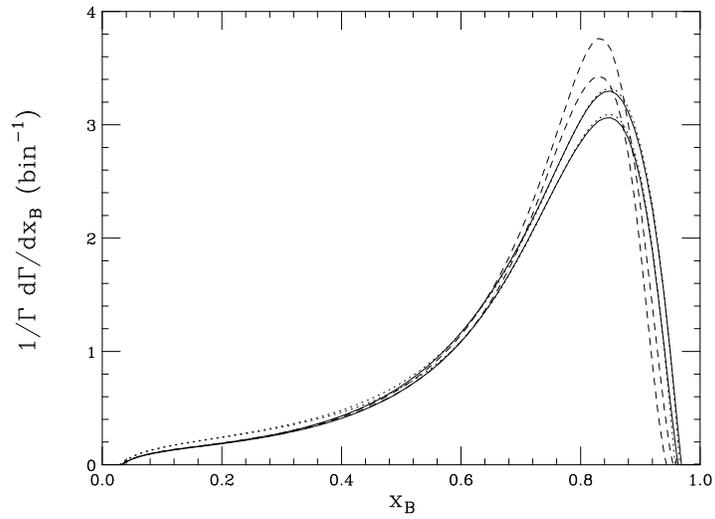}}} 
\caption{As in Fig.~\ref{had1}, using the model of
Eq.~(\ref{ab}) and fitting 
its parameters to ALEPH (solid lines) and SLD (dashes), including NLL soft
resummation in the initial condition of the 
perturbative fragmentation function. We also show results from the fit 
to ALEPH, but with no soft resummation (dots).}
\label{had3}
\end{figure}

In Fig.~\ref{had3}  we compare the ALEPH and SLD predictions
according to the
power law of Eq.~(\ref{ab}) since,
as shown in Table~\ref{tableres}, this is the only hadronization model
where the fitted 
parameters are statistically different. In fact, the overall 
shapes of the ALEPH- and 
SLD-based distributions lead to different predictions 
within one-standard-deviation
accuracy, expecially for $x_B\gsim 0.7$, with the SLD ones being peaked 
at smaller $x_B$ values. This result can be checked to be true even at higher 
confidence level and, as anticipated, can be associated with the different 
$b$-hadron samples which the two experiments reconstructed. 
For the sake of comparison, we also show the ALEPH-based
prediction without resumming soft logarithms, as Table~\ref{tablenores}
reports a small $\chi^2$ even in this case. We find that, although 
the perturbative predictions look rather different according to whether
such contributions are resummed or not (see
Fig.~\ref{figpart}), the hadron-level results agree for $x_B\gsim 0.5$.
In fact, the convolution with the non-perturbative fragmentation
function smears the sharp peak shown by the unresummed result and
the parameters $\alpha$ and $\beta$ are accordingly set by the fit to
the $e^+e^-$ data in such a way that the two $x_B$-predictions do not differ
too much from each other.
Statistically-significant differences are nonetheless found
for $x_B\lsim 0.5$, where the predicted bands get narrower.  
\begin{figure}
\centerline{\resizebox{0.60\textwidth}{!}{\includegraphics{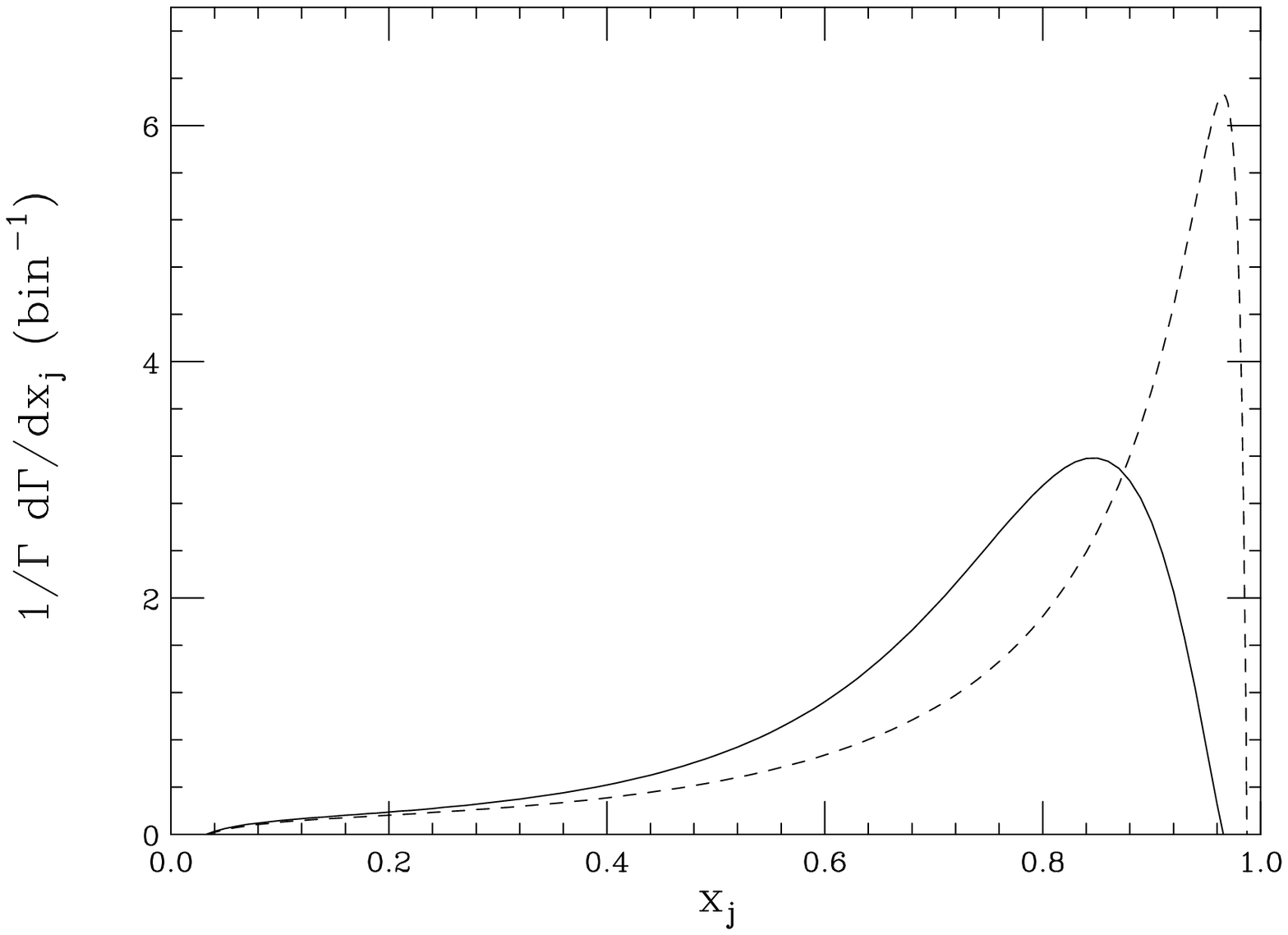}}} 
\caption{Comparison of hadron- ($j=B$, solid) and parton-level 
($j=b$, dashes) results using the hadronization model of Eq.~(\ref{ab}), 
with $\alpha$ and $\beta$ being the central values in the fit 
to the ALEPH data, and including NLL soft resummation in the initial condition
of the perturbative fragmentation function.}
\label{had4}
\end{figure}
\par
It is finally interesting to gauge the overall impact of the non-perturbative 
fragmentation by comparing our parton-level result with one of our
hadron-level predictions, in particular the one obtained using the power law
with two parameters fitted to the ALEPH data, as it corresponds 
to our smallest $\chi^2$. From Fig.~\ref{had4},
we learn that the hadronization effects are remarkable and 
the $x_B$ distribution is considerably softened with respect to the
$x_b$ one. This means that, even when one uses a refined perturbative
approach, with DGLAP evolution and soft resummation in the initial condition
of the perturbative fragmentation function up to NLL accuracy,
the role played by the
non-perturbative input and hence by the $e^+e^-$ experimental data 
will still be 
crucial to perform any prediction on $b$-fragmentation in top decay.
\section{Conclusions}
We discussed the $b$-quark fragmentation in top decay in NLO QCD using
the method of perturbative fragmentation, 
which resums large logarithms $\sim\log(m_t^2/m_b^2)$ 
which multiply the strong  
coupling constant in the fixed-order massive calculation.

We computed the NLO differential width of top decay with respect to the
$b$-quark energy fraction $x_b$
for a massless $b$, fully including $b$-mass effects and for a massive
$b$ quark, but neglecting contributions proportional to powers of the ratio
$m_b/m_t$. We determined the $\overline{\mathrm{MS}}$-subtracted coefficient 
function and checked that our result is consistent with the known 
expression of the initial condition for heavy-quark
perturbative fragmentation functions.
We convoluted our $\overline{\mathrm{MS}}$ coefficient function with
the process-independent, 
perturbative fragmentation function for a massive $b$ quark,
evolved up to NLL accuracy using the DGLAP equations.
We showed results for the $b$ energy-fraction distribution
in top decay, which we compared to the fixed-order results for 
a massive $b$ quark. 
We found that the use of the perturbative fragmentation approach has
a remarkable effect on the parton-level distribution which is smoothed
out with respect to the ${\cal O} (\alpha_S)$ one, which gets arbitrarily
large once $x_b$ approaches unity.
We also investigated the impact of the resummation of 
process-independent next-to-leading  
logarithms, which appear in the initial condition of the
perturbative fragmentation function $D_b(x_b,\mu_{0F},m_b)$ and are 
associated with soft-gluon radiation. 
We found that it
softens the $x_b$ distribution and makes the dependence on the
scale $\mu_{0F}$ weaker.

We then studied the energy distribution of $b$-hadrons in top 
decay and fitted some hadronization models to    
$e^+e^-$ data. We used ALEPH data on  
$b$-flavoured mesons and SLD data on $b$-flavoured baryons and mesons. 
In order to perform such fits we described the perturbative process 
$e^+e^-\to b\bar b (g)$ as we did for $b$-quark production in 
top decay.
Throughout our analysis, we discarded data points where the hadron-level
energy fraction is close to $x_B\simeq 0,1$, 
since our approach is unreliable in the neighbourhood of these points.

We found that, within our perturbative framework, 
models which describe the non-perturbative fragmentation 
according to power laws with one and, in particular, with
two fittable parameters lead to  
good descriptions of ALEPH and SLD data.
The Peterson model is marginally consistent with the ALEPH
results and unable to describe the SLD data. 
We also found that, although the ALEPH and SLD data refer to different 
$b$-hadron samples, the best-fit parameters obtained using the Kartvelishvili
or the Peterson model are statistically-consistent within the error ranges.  
The implementation of NLL soft-gluon resummation in the initial
condition of the perturbative
fragmentation function yields pretty different fits to the $e^+e^-$ data.
An unresummed perturbative
calculation still leads to good fits to the ALEPH data once we use
power laws to model the hadronization, while it is unable to reliably 
describe the SLD data. 

We then showed distributions of the energy fraction $x_B$ of 
$b$-flavoured hadrons in top decay, using
only models which give reliable descriptions of the $e^+e^-$ data. 
We found that the models fitted to the ALEPH data yield  
statistically-distinguishable predictions
for the $b$-meson spectrum in top decay. In particular,
the Peterson-model distributions lie quite far from the others 
and are peaked at slightly-larger $x_B$ values.  
If we model the hadronization using 
power laws with one or two parameters, fitted to the 
SLD data sample, the predictions for the $x_B$ spectrum in top decay are 
compatible within two standard deviations, a result which is due
to the large uncertainties on the best-fit values of $\alpha$ and 
$\beta$. 
We also compared results obtained using the power law with two parameters,
but fitted to the ALEPH or SLD data, and obtained predictions which
are statistically different.

We investigated the impact of NLL soft-gluon resummation in the 
initial condition of the perturbative
fragmentation function on hadron-level distributions using ALEPH-based fits. 
We found a significant impact on the
$x_B$ spectrum only at relatively-small $x_B$, while  
predictions with or without soft resummation are indistinguishable
at large $x_B$ values.
This result stresses the importance of
the non-perturbative input and consequently of the use of the $e^+e^-$ results
to perform reliable predictions in top decay. This can be 
learned from direct comparison between parton- and hadron-level 
distributions shown throughout the paper.

It will be now very interesting to use the present approach
to perform predictions of other 
observables relying on the $b$-fragmentation in top decay, such
as the invariant-mass distributions used in [\ref{avto},\ref{cmlms}]
to fit the top mass value.
It is clearly mandatory to compare the results obtained in the framework
of perturbative fragmentation functions with the ones of Monte Carlo
event generators, taking particular care about 
the induced uncertainty on the top mass measurement. 
However, for such a comparison to be
reliable, Monte Carlo programs will have to be tuned to fit with
the experimental data. 
This is in progress.       

Finally, we plan to extend the method developed in [\ref{cc}] 
in the framework of $e^+e^-$ processes to resum with NLL accuracy
also the process-dependent 
soft logarithms $\sim\alpha_S^n(\mu)\log^{n+1}N$ and 
$\sim\alpha_S^n(\mu)\log^nN$ 
appearing in the Mellin transform of the $\overline{\mathrm{MS}}$
coefficient function (\ref{diff}). 
This further step will allow one to include in the
perturbative calculation all soft logarithms, besides the 
process-independent ones which are present in the initial condition of the 
perturbative fragmentation function and have been 
correctly accounted for throughout this paper. 
It will be interesting to analyse the impact of soft-gluon resummation
in the $\overline{\mathrm{MS}}$ coefficient function on the energy
spectrum of $b$ quarks and $b$-flavoured hadrons.
This is in progress as well. 
\section*{Acknowledgements}
We are especially indebted to M. Cacciari for useful correspondence and
for providing us with the computing code to perform numerical inversions 
of Mellin-space formulas and fits of hadronization models to 
$e^+e^-$ data. We acknowledge T. Boccali, D. Dong,
R.H. Hemingway, L.H. Orr, C. Macesanu, K.S. McFarland,
C. Oleari, M.H. Seymour and D. Wackeroth for discussions
on these and related topics.
The work of G.C. was supported  
by grant number DE-FG02-91ER40685 from the U.S.
Department of Energy.

\begin{appendix}
\section{Details of the calculation}
We wish to give some details on the calculation of the top-decay 
differential width at ${\cal O(\alpha_S)}$ for massless and massive
$b$ quarks. In particular, the comparison of the two results will allow us to
obtain the initial condition (\ref{dbb}) of the
perturbative fragmentation function.
We adopt the on-shell mass-renormalization scheme and
use dimensional regularization to regulate the ultraviolet and 
soft singularities and,
in the massless case, the collinear divergence as well. 
We define the parameter $\epsilon$, which is   
related to the number $d$ of dimensions via $d=4-2\epsilon$.

In the massless case, the differential width will cointain a pole 
$\sim 1/\epsilon$, due to the collinear singularity, which disappears only 
in the total NLO width.  
This requires that, in order to get the correct finite term in the  
normalized $(1/\Gamma_0)d\hat\Gamma_b/dx_b$, with $x_b$ defined 
in Eq.~(\ref{xbnorm}), 
the Born width $\Gamma_0$ will have to be evaluated in dimensional
regularization at ${\cal O} (\epsilon)$.
We find:
\begin{equation}
\Gamma_0={{\alpha m_t|V_{tb}|^2}\over{16\sin^2\theta_W}}
{{(1-w)^2(1+2w)}\over w}\left\{1+\epsilon\left[-\gamma_E+\log 4\pi-
2\log (1-w) +2 {{1+w}\over {1+2w}}\right]\right\}, 
\end{equation}
where $\alpha$ is the electromagnetic coupling constant, 
$\gamma_E=0.577216\dots$ is the Euler constant, $\theta_W$ is
the Weinberg angle and 
$w$ is the ratio $w=m_W^2/m_t^2$, already introduced in Section 2.
We thus obtain:
\begin{eqnarray}
{1\over{\Gamma_0}}  {{d\hat\Gamma_b}\over {dx_b}}&=&
\delta(1-x_b)+{{\alpha_S(\mu)}\over{2\pi}}\left\{C_F
\left[ {{1+x_b^2}\over{(1-x_b)_+}}+{3\over 2}\delta (1-x_b)\right]\right.
\nonumber\\
&\times&\left(-{1\over\epsilon}+\gamma_E-\log 4\pi\right)
+\hat A_1(x_b)\Bigg\},
\label{cmseps}
\end{eqnarray}
with $\hat A_1(x_b)$ defined in Eq.~(\ref{cms}). 
We note that $\hat A_1(x_b)$ depends on
the scale $\mu_F$, remnant of the regularization procedure, which 
disappears only in the total width.
We also checked that the integral of
Eq.~(\ref{cmseps}) agrees with the result of [\ref{wid1}], where 
the ${\cal O}(\alpha_S)$ corrections to the top width have been
evaluated in the approximation of a massless $b$ quark.
In order to get the $\overline{\mathrm{MS}}$ coefficient function 
(\ref{diff}), 
by definiteness, we shall have to subtract from Eq.~(\ref{cmseps}) the
${\cal O} (\alpha_S)$ term multiplying the
characteristic $\overline{\mathrm{MS}}$ constant 
$(-1/\epsilon +\gamma_E-\log 4\pi)$.

We wish to derive the NLO differential width with the full inclusion  
of the $b$ mass $m_b$. For this purpose, it is convenient to define
the following quantities:
\begin{eqnarray}
b&=& {{m_b^2}\over {m_t^2}},\\
s&=&{1\over 2} (1+b-w),\\
\beta &=& {\sqrt{b}\over s},\\
Q&=& s\sqrt{1-\beta^2},\\
G_0&=&{1\over 2}\left[ 1+b-2w+{{(1-b)^2}\over w}\right],\\
\Phi (x_b)&=&s\left[\sqrt{x_b^2-\beta^2}-
\mathrm{arcth} \left( {{\sqrt{x_b^2-\beta^2}}\over {x_b}}\right)\right],
\end{eqnarray}
where $x_b$ is the normalized $b$-quark energy fraction
\begin{equation}
x_b={{x_E}\over {x_{E,\mathrm{max}}}}={{x_E}\over 2s}
\ \ \ , \ \ \ \beta\leq x_b \leq 1,
\end{equation}
with $x_E$ defined in Eq.~(\ref{xbpart}).
We find:
\begin{eqnarray} 
{1\over{\Gamma_0}}{{d\Gamma}\over{dx_b}}&=&\delta(1-x_b)+ 
{{C_F\alpha_S(\mu)}\over{\pi Q}}\left\{\left\{2s\left[
{\mathrm{Li}}_2\left( {{2Q}\over{1-s+Q}}\right)-
{\mathrm{Li}}_2\left( {{2Q}\over{s-b+Q}}\right)\right.\right.\right.
\nonumber\\
&-&\log(s+Q)\left(\log\left({{1-s+Q}\over {\sqrt{w}}}\right)+
\log\left( {{s-b+Q}\over{2s(1-\beta)}}\right)\right)\nonumber\\
&+&\left.{1\over 2}
\log (b)\log \left({{s-b+Q}\over{2s(1-\beta)}}\right) \right]+
\left( 3{{Q^2}\over{G_0}}+s-b\right)\log\left({{s+Q}\over \sqrt{b}}\right)
\nonumber\\
&+&(1-b)\log\left({{1-s+Q}\over {\sqrt{w}}}\right)+ Q
\left[\left(6{{(w-b)(s-b)}\over{wG_0}}-1\right){{\log (b)}\over 4}
\right.\label{fullmass}\\
&-&\left.\left.2\ \log\left({{2s(1-\beta)}\over{\sqrt{w}}}\right)-2
\right]\right\}\delta(1-x_b)\nonumber\\
&-&2\ {\Phi (x_b)}\left[ {1\over {(1-x_b)_+}}+
{s\over{G_0}}\left(1+{{1+b}\over{2w}}\right)(1-x_b) - 1\right]\nonumber\\
&+&\left. 2s\sqrt{x_b^2-\beta^2}\left[ 
2{s^2\over G_0}\left( {{1-x_b}\over{1-2sx_b+b}}\right) + 
{s\over{G_0}}\left(1+{{1+b}
\over{2w}}\right)(1-x_b) - 1\right]\right\}\nonumber.
\end{eqnarray}
In Eqs.~(\ref{cmseps}) and (\ref{fullmass})
we note the presence of the so-called
`plus prescription' $1/(1-x_b)_+$. Such a term arises after one integrates
over the phase space for real-gluon radiation. 
For a massive $b$ quark, where $x_{b,\mathrm{min}}\neq 0$, one
makes use of the following expansion:
\begin{equation}
{{(x_b-x_{b,\mathrm{min}})^\epsilon}\over{(1-x_b)^{1+\epsilon}}}=
-{1\over\epsilon}\delta(1-x_b)+{1\over{(1-x_b)_+}}+{\cal O}(\epsilon),
\label{plus}
\end{equation}
with the plus prescription defined as:
\begin{equation}
\int_{x_{b,\mathrm{min}}}^1{\left(g(x_b)\right)_+ h(x_b)dx_b}=
\int_{x_{b,\mathrm{min}}}^1{g(x_b) [h(x_b)-h(1)]dx_b}.
\label{plusint}
\end{equation}

We observe that in Eq.~(30) the quark mass regulates the collinear 
divergence, therefore, 
unlike the result with a massless $b$ quark 
(\ref{cmseps}), Eq.~(30) does not contain any dependence on
the dimensional-regularization quantities $\epsilon$ or $\mu_F$.

For the sake of checking the initial condition of the $b$-quark perturbative
fragmentation function (\ref{dbb}), we need to rewrite 
Eq.~(30) neglecting powers of $m_b/m_t$.

We find:
\begin{equation}
{1\over{\Gamma_0}}{{d\Gamma}\over {dx_b}}=\delta(1-x_b)
+{{\alpha_S(\mu)}\over{2\pi}}A_1(x_b),
\label{gammass}
\end{equation}
with
\begin{eqnarray}
A_1(x_b)&=&C_F\ \left\{ \left[ {{1+x_b^2}\over{(1-x_b)_+}}
+{3\over 2}\delta(1-x_b)\right]\log{{m_t^2}\over{m_b^2}}\right.\nonumber\\
&+& 2{{1+x_b^2}\over {(1-x_b)_+}}\log[(1-w)x_b]-{{4x_b}\over{(1-x_b)_+}}
+{{4w(1-w)}\over{1+2w}}{{x_b(1-x_b)}\over {1-(1-w)x_b}}\nonumber\\
&+& \delta(1-x_b)\left[ 4{\mathrm{Li}}_2 (1-w)+2\log w\log (1-w)
-{{2\pi^2}\over 3}-{{2(1-w)}\over{1+2w}}\log(1-w)\right.\nonumber\\
&-&{{2w}\over {1-w}}\log w -4\Bigg]\Bigg\}.
\label{mass}
\end{eqnarray}
We note that in Eq.~(\ref{mass}) the strong coupling constant
multiplies the large logarithm
$\log(m_t^2/m_b^2)$, which spoils the reliability of the fixed-order
calculation and makes the approach of perturbative
fragmentation mandatory. 
Such large logarithms are absent in the total NLO width, which 
can be checked to agree
with the result in the literature, once one accounts for the mass 
of the $b$ quark [\ref{wid2}]. 

If $\mu_F$ is of the order of $m_b$, 
one can express the perturbative fragmentation function
$D(x_b,\mu_F,m_b)$ as a power expansion in $\alpha_S$:
\begin{equation}
D(x_b,\mu_F,m_b)=d^{(0)}(x_b)+
{{\alpha_S(\mu)}\over{2\pi}}d^{(1)}(x_b,\mu_F,m_b)
+{\cal O}(\alpha_S^2)
\label{d01}
\end{equation}
Inserting Eqs.~(\ref{diff}), 
(\ref{gammass}) and (\ref{d01}) in Eq.~(\ref{pff}), and
solving for $d^{(0)}$ and $d^{(1)}$, one finds:
\begin{eqnarray}
d^{(0)}(x_b)&=&\delta(1-x_b),\\
d^{(1)}(x_b,\mu_F,m_b)&=&A_1(x_b)-\hat A_1(x_b).
\end{eqnarray}
Comparing then $A_1(x_b)$ and $\hat A_1(x_b)$, it is straightforward 
getting Eq.~(\ref{dbb}). It should be noted that, although $A_1(x_b)$  
and $\hat A_1(x_b)$ separately depend on $m_W$ via the ratio $w$, their 
difference and hence the initial condition for the perturbative fragmentation
function do not, which is essential for it to be process independent.
\section{Coefficient function in Mellin space} 
For the sake of completeness, we wish to present the result for the Mellin
transform of the  $\overline{\mathrm{MS}}$ coefficient function (\ref{diff}):
\begin{equation}
\hat\Gamma_N (m_t,m_W,\mu,\mu_F)=
{1\over{\Gamma_0}}\int_0^1 {dz \ z^{N-1}
{{d\hat\Gamma_b}\over{dz}}^{\overline{\mathrm{MS}}}
(z,m_t,m_W,\mu,\mu_F) }.
\end{equation}
Given $\Gamma(x)$ the Euler Gamma function, we define the poligamma functions:
\begin{equation}
\psi_k(x)={{d^k\log\Gamma(x)}\over {dx^k}}
\end{equation} 
and the combinations
\begin{eqnarray}
S_1(N)&=&\psi_0(N+1)-\psi_0(1),\\
S_2(N)&=&-\psi_1(N+1)+\psi_1(1).
\end{eqnarray}
The basic, non-trivial $N$-space transforms of the terms which 
appear in 
Eq.~(\ref{diff}) are given by:
\begin{eqnarray}
\int_0^1{dz{{z^{N-1}}\over{(1-z)_+}}}&=&-S_1(N-1),\\
\int_0^1{dz{{\log z}\over{(1-z)_+}}z^{N-1}}&=&-\psi_1(N),\\
\int_0^1{dz\left[{1\over{1-z}}\log(1-z)\right]_+z^{N-1}}&=&{1\over 2} 
\left[S_1^2(N-1)+S_2(N-1)\right],\label{s1s2}\\
\int_0^1{dz{{z(1-z)}\over{1-(1-w)z}}z^{N-1}}&=&
{{_2F_1(1,N+1,N+2,1-w)}\over{N+1}}\nonumber\\
&-&{{_2F_1(1,N+2,N+3,1-w)}\over{N+2}}.
\end{eqnarray}
We shall then get:
\begin{eqnarray}
\hat\Gamma_N (m_t,m_W,\mu,\mu_F)&=&1+{{\alpha_S(\mu)C_F}\over{2\pi}}
\left\{\log{{m_t^2}\over{\mu_F^2}}\left[{1\over{N(N+1)}}-2S_1(N)+
{3\over 2}\right]\right.\nonumber\\
&+&[1+2\log(1-w)]{1\over{N(N+1)}}-2\psi_1(N)
-2\psi_1(N+2)\nonumber\\
&+&{{4w(1-w)}\over{1+2w}}
\left[{{_2F_1(1,N+1,N+2,1-w)}\over{N+1}}\right.\nonumber\\
&-&\left. {{_2F_1(1,N+2,N+3,1-w)}\over{N+2}}\right]
+S_1^2(N-1)+S_1^2(N+1)\nonumber\\
&+&S_2(N-1)+S_2(N+1)+2[1-2\log(1-w)]S_1(N)\nonumber\\
&+&2\log w\log(1-w)-2{{1-w}\over{1+2w}}\log(1-w)
-{{2w}\over{1-w}}\log w\nonumber\\
&+&4{\mathrm{Li}}_2(1-w)-6-{{2\pi^2}\over 3}\Bigg\}.
\end{eqnarray}
One can show that, for $N\to \infty$:
\begin{equation}
S_1(N)\sim \psi_0 (N)\sim \log N.
\end{equation}
Similar large-$N$ behaviour is also shown by the Mellin transform 
of the initial condition of the perturbative fragmentation function
(\ref{dbb}), whose most-singular term at large $x_b$ has a $N$-space
counterpart analogous to Eq.~(\ref{s1s2}). 
\end{appendix}
\section*{References}
\begin{enumerate}
\addtolength{\itemsep}{-0.5ex}
\item\label{lhc}
M. Beneke, I. Efthymiopoulos, M.L. Mangano, J. Womersley et al., 
in Proceedings of 1999 CERN Workshop on
Standard Model Physics (and more) at the LHC, CERN 2000-004, 
G. Altarelli and M.L. Mangano eds., p.~419, hep-ph/0003033.
\item\label{lc}
J.A. Aguilar--Saavedra et al., ECFA/DESY LC Physics Working group,
hep-ph/0106315;\\
T. Abe et al., American Linear Collider Working Group, Part 3, Studies
of Exotic and Standard Model Physics, hep-ex/0106057; \\
K. Abe et al., ACFA Linear Collider Working group, hep-ph/0109166.
\item\label{tev}
D\O\ Collaboration, B. Abbott et al., Phys.\ Rev.\ D58 (1998) 052001;\\ 
CDF Collaboration, T. Affolder et al., Phys.\ Rev.\ D63 (2001) 032003.
\item\label{avto}
A. Kharchilava, Phys.\ Lett.\ B476 (2000) 73.
\item\label{herwig}
G. Corcella, I.G. Knowles, G. Marchesini, S. Moretti, K. Odagiri,  
P. Richardson, M.H. Seymour and B.R. Webber, JHEP 0101 (2001) 010.  
\item\label{pythia}
T. Sj\"ostrand, Comput.\ Phys.\ Commun.\ 82 (1994) 74;\\
T. Sj\"ostrand, P. Ed\'en, C. Friberg, 
L. L\"onnblad, G. Miu, S. Mrenna and E. Norrbin, 
Comput.\ Phys.\ Commun.\ 135 (2001) 238.
\item\label{isajet}
H. Baer, F.E. Paige, S.D. Protopopescu and X. Tata, 
hep-ph/0001086.
\item\label{webber}
B.R. Webber, Nucl.\ Phys.\ B238 (1984) 492.
\item\label{string}
B. Andersson, G. Gustafson, G. Ingelman and T. Sj\"ostrand, Phys.\ Rep.\ 97 
(1983) 31. 
\item\label{feyn}
R.D. Field and R.P. Feynman, Nucl.\ Phys.\ B136 (1978) 1.   
\item\label{marweb}
G. Marchesini and B.R. Webber, Nucl.\ Phys.\ B310 (1988) 461.
\item\label{mecorr}
G. Corcella and M.H. Seymour, Phys.\ Lett.\ B442 (1998) 417;\\
E. Norrbin and T. Sj\"ostrand, Nucl.\ Phys.\ B603 (2001) 297.  
\item\label{cmlms}
G. Corcella, J.\ Phys.\ G26 (2000) 634;\\
G. Corcella, M.L. Mangano and M.H. Seymour, JHEP 0007 (2000) 004. 
\item\label{collins}
J.C. Collins and D.E. Soper, Ann.\ Rev.\ Nucl.\ Part.\ Sci.\ 37 (1987) 383.
\item\label{mele}
B. Mele and P. Nason, Nucl.\ Phys.\ B361 (1991) 626.
\item\label{ap}
G. Altarelli and G. Parisi, Nucl.\ Phys.\ B126 (1977) 298.
\item\label{dgl}
L.N. Lipatov, Sov.\ J.\ Nucl.\ Phys.\ 20 (1975) 95;
V.N. Gribov and L.N. Lipatov, Sov.\ J.\ Nucl.\ Phys.\ 15 (1972) 438;
Yu.L. Dokshitzer, Sov.\ Phys.\ 46 (1977) 641.
\item\label{cc}
M. Cacciari and S. Catani, hep-ph/0107138.
\item\label{cagre1}
M. Cacciari and M. Greco, Nucl.\ Phys. B421 (1994) 530.
\item\label{kart}
V.G. Kartvelishvili, A.K. Likehoded and V.A. Petrov, Phys.\ Lett.\ B78 (1978)
615. 
\item\label{peterson}
C. Peterson, D. Schlatter, I. Schmitt and P.M. Zerwas, Phys.\ Rev.\ D27 (1983)
105.
\item\label{colnas}
G. Colangelo and P. Nason, Phys.\ Lett.\ B285 (1992) 167.
\item\label{nasweb}
P. Nason and B.R. Webber,  
Nucl.\ Phys.\ B421 (1994) 473; ibid. B480 (1996) 755 (erratum). 
\item\label{cagre2}
M. Cacciari and M. Greco, Phys.\ Rev.\ D55 (1997) 7134. 
\item\label{nasole}
P. Nason and C. Oleari, Nucl.\ Phys.\ B565 (2000) 245. 
\item\label{cagre3}
M. Cacciari, M. Greco, S. Rolli and A. Tanzini, Phys. Rev. D55 (1997) 2736.
\item\label{cagre4}
M. Cacciari and M. Greco, Z.\ Phys.\ C69 (1996) 459. 
\item\label{vtb}
CDF Collaboration, T. Affolder et al., Phys.\ Rev.\ Lett.\ 86 (2001) 3233.  
\item\label{pij}
G. Curci, W. Furmanski and R. Petronzio, Nucl.\ Phys.\ B175 (1980) 27;\\
W. Furmanski and R. Petronzio, Phys.\ Lett.\ B97 (1980) 437.
\item\label{pij1}
E.G. Floratos, D.A. Ross and C.T. Sachrajda, Nucl.\ Phys.\ B129 (1977) 66; 
ibid. B139 (1978) 545 (erratum); ibid. B152 (1979) 493;\\
A. Gonzales--Arroyo, C. Lopez and F.J. Yndurain, Nucl.\ Phys.\ B153 (1979) 
161;\\
E.G. Floratos, R. Lacaze and C. Kounnas, Nucl.\ Phys.\ B192 (1981) 417.
\item\label{wid1}
C.S. Li, R.J. Oakes and T.C. Yuan, Phys. Rev. D43 (1991) 855.
\item\label{spira}
B.A. Kniehl, G. Kramer and M. Spira, Z.\ Phys.\ C76 (1997) 689.
\item\label{aleph}
ALEPH Collaboration, A. Heister et al., Phys.\ Lett.\ B512 (2001) 30.
\item\label{sld}
SLD Collaboration, K. Abe et al., Phys.\ Rev\ Lett. 84 (2000) 4300;\\
D. Dong, private communication.
\item\label{caccia}
M. Cacciari, private communication.
\item\label{wid2}
A. Czarnecki, Phys.\ Lett.\ B252 (1990) 467.
\end{enumerate}
\end{document}